\pdfoutput=1
\documentclass[twocolumn]{article}
\usepackage{graphicx}
\usepackage{grffile}

\begin{document}
\title{Comparitive Study of Geometric and Image Based Modelling and Rendering Techniques.}
\author{
  Seth, Agrima\\
  Army Institute of Technology,\\
  University of Pune,\\
  Pune,\\ 
  India,\\
  \texttt{agrimaseth\_12528@aitpune.edu.in}
  \and
  Mishra,Dr.Deepak\\
  Indian Institute of Space Science and Technology\\
  Trivandrum\\
  India\\
  \texttt{deepak.mishra@iist.ac.in}
}
\maketitle

\begin{abstract}
This is a comparative study of the traditional 3D computer graphics technique of geometric modelling and image-based rendering techniques that were surveyed and implemented.We have discussed the classifications and representative methods of both the techniques. The study has shown that there is a strong continuum between both the techniques and a hybrid of the two is most suitable for further implementations.This hybridisation study is underway to create models of real life situations and provide disaster management training.
\end{abstract}

\section{Introduction}
The expeditious development in processor technology and new levels of embedded designs have made computers with faster and more efficient graphic boards. These have inturn made it possible to move(virtually) into a new world, of both reality and imagination. It allows one to experience the unventured and undiscovered world.
\par Ever since the inception of this field the desire of the users-industrial and scientific alike,have grown.The seed of this idea was sown by Ivan Sutherland in 1965 "make that virtual world in the window look real,sound real,feel real and respond realistically to viewes actions"[1].Thus the essential idea of this paper is to find a method for maximising the reality of the worlds and the user interactions with the world.
\subsection{What is ideal modelling?}
\begin{itemize}
\item "Virtual reality refers to immersive,interactive and computer generated environments and the combination of technologies required to build these environments"[2]

\item "Virtual reality lets you navigate and view a world of three dimensions in real time with six degrees of freedom"[3] 

\end{itemize}
In this paper we shall focus on comparison of technologies required to bulid desktop virtual reality, and use of datagloves to interact with the environment

\section{Geometric Modelling}
We use mathematical models to model the realistic objects of computer graphics.
The method stores the information of the object as geometric elements(lines,polygons,polyhedrons, etc).

Representation of an object in the mathematical space using a set of points is termed as modelling and its description using information models is representation scheme.
Here the tool under consideration is 3Ds Max studio.

\subsection{Modelling Method}
In geometric modelling predetermining the structure,shape and texture of the model to be rendered is essential.They are created by using curve and point transformations for each polygon,edge and vertex(as per the requirement).Hence, capturing the spatial aspects of objects for an application.

\textbf{\textit{•Definition 1:}}Geometric information is called a set of attribute classes, which is perceived by one as a (rigid) solid. We denote it by the ordered triplet:

G = ({s}, {m}, {p})
 
where {s} is the set of spatial forms, {m} is the set of metric characteristics, and {p} is the set of parameters giving the location and
orientation.

Each geometric information G induces in the mathematical space (for example in \[E^3\]) an abstract object (set of points) or a family of abstract objects defined by the numerical characteristics of {m} and {p}, as well as the form {s}.
Obviously, two different pieces of geometric information G1 and G2 may describe the
same sets of points. Hence the problem of comparing G1 and G2 arises. Two of the main tasks of geometric modelling are finding and generating equivalent and identical pieces of geometric information. [4]

\subsection{Representation Scheme}
3Ds Max Studio uses the representaion schemes discussed below:

\begin{itemize}
\item Boundary Representation Scheme
\item Constructive solid Geometry Scheme
\item Parameter and feature Based Scheme
\item Sweeping Scheme
\item Particle System Scheme
\end{itemize}

\textit{B-Rep:}To visualise the object it uses the boundary description,\textit{i.e;}of their edges; which inturn is described by their end points.An analogous representation is used for faces(NURBS surfaces).

It allows a quick,high quality and realistic visualisation of the object space. It uses the ray-tracing algorithm for more realistic visualisation and hence, creating the effects of reflections such as specular,ambient and others. Whereas, for its faster and quick implementation z-buffer algorithm is applied[4][5][6].

\textit{CSG:}It represents solids as boolean constructions and is used in combination with B-Rep due to its simple data structure and elegant recursive method to represent solids[4][5].

\textit{Sweeping Scheme:}It is used to describe rotary solids and removal of material, as they represent solids as a parametrisation of primitives in space.

\textit{Particle System Scheme:}It is used to create highly complex special effects such as snow,fire,clouds. It uses addition of  thousands(and millions)of patricles and also their evolution in terms of space and time to create a given effect[5].

\section{Image Based Rendering using Explicit Geometry}
In the following technique,we have used the 3D information encoded in the depth;along the line of sight.
\subsection{3-D Warping}
When the depth information of all the points to be rendered is available through either of the input images.To render images the pixels of the original image are projected to their respective 3-D locations and re-projected to the new image.

\subsubsection{Improving speed}The rendering speed decreases with increase in complexity of the images.Hence, to effectively handle it relief-texture technique by oliveira and bishop[7].

\subsubsection{Image perspective} The images used for the implementation of 3-D warping technique were multi-perspective.The final view was achieved by warping all the muti-perspective images[8].

\subsection{Occlusion Handling}
The major disadvatage of this technique is the occlusions generated in the output image. These occlusions are generated due to difference of sampling resolution between the input and output images.To render these holes with the image pixel splatting is applied.Another approach could be use of layered depth images[9] storing both what is visible in the input image and also what is not visible. The second approach proved more effective in implementaion.

\subsection{Texture Mapping}
To achieve photo-realistic environment, texture needs to be mapped onto the synthetic models.To achieve maximum accuracy for effects such as reflections,transparency a three step view-dependent texture mapping proposed by debevec \textit{et al.}is used. It uses visibility pre-processing,polygon-view maps and projective texture mapping. The three step processing makes the blend faster and smoother[10].

\subsection{Multiple View Point Rendering}
It bridges the idea of lumnigraph and 3D scene geometry.This technique uses the already known 3D scene to compute viewpoints from known camera viewpoints and hence utilises the perspective coherence

The following figure depicts the result generated:
\begin{center}

\textit{a}.The disparity map generated
\end{center} 
\includegraphics[scale=0.2]{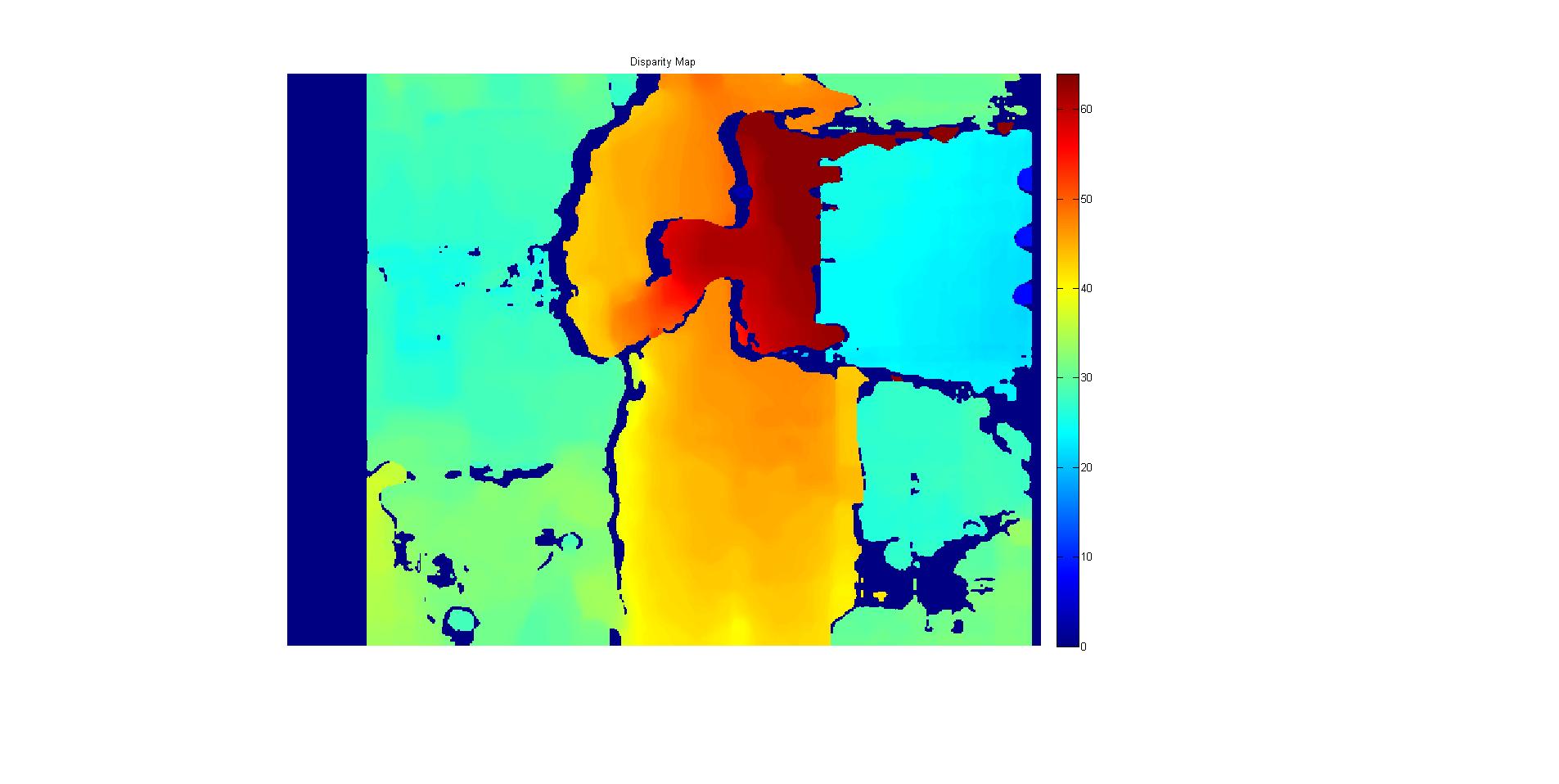}
 
 \begin{center}
•\textit{b}.\newpage The cloud image generated
\end{center}
\hspace*{-3cm}\includegraphics[scale=0.2]{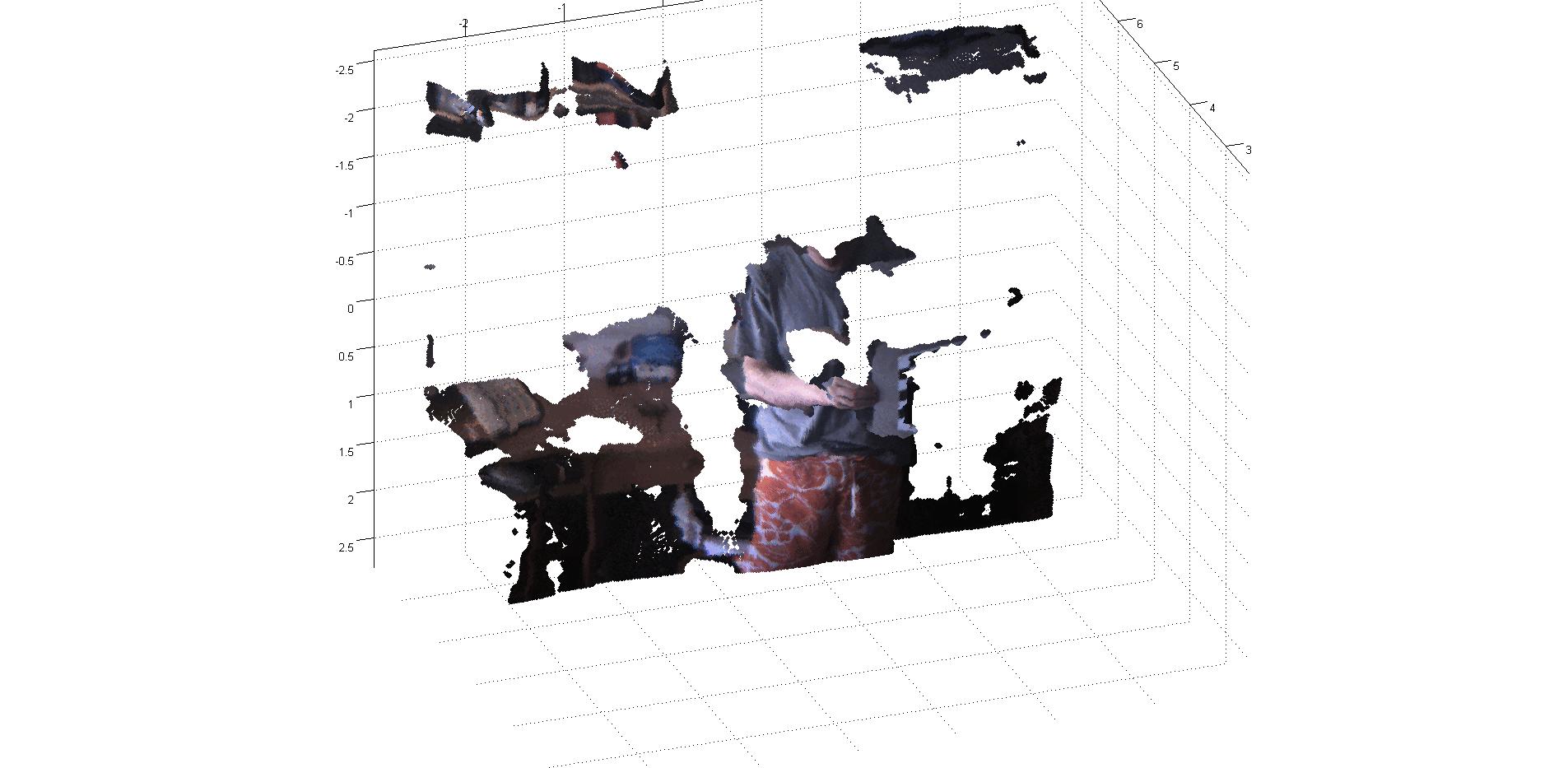}

\section{Conclusion}
A comparison of traditional 3D computer graphics technique of geometric modelling and image-based rendering techniques was done. The geometrical representation provides the advantage of being compressive;but lacks in texture dynamicity. On the other hand, image based rendering have photorealistic rendering but it comprises of extensive data acquisition and immense storage requirement.

Thus, we conclude that since these are both complimentary to each other, it is essential to form a hybrid of the two and hardware be designed for a collective implementation of both.

\end{document}